\documentclass[conference]{IEEEtran}
\IEEEoverridecommandlockouts
\usepackage{cite}
\usepackage{amsmath,amssymb,amsfonts}
\usepackage{algorithmic}
\usepackage{graphicx}
\usepackage{textcomp}
\usepackage{xcolor}

\usepackage{booktabs}
\usepackage{tikz}

\usepackage{comment}
\usepackage[utf8]{inputenc}
\usepackage{soul}
\usepackage{makecell}

\usepackage{tabularx}
\usepackage{soul}

\def\BibTeX{{\rm B\kern-.05em{\sc i\kern-.025em b}\kern-.08em
    T\kern-.1667em\lower.7ex\hbox{E}\kern-.125emX}}
\begin{document}

\title{QUIC-TRIP: A Triple-Redundant Journey Toward Secure Substation Communications\\ \vspace{-2mm}
}

\author{\IEEEauthorblockN{Jorge David de Hoz Diego}
\IEEEauthorblockA{\textit{School of Computer Science} \\
\textit{University College Dublin}\\
jorge.dehozdiego@ucd.ie}
\and
\IEEEauthorblockN{Ioannis Zografopoulos}
\IEEEauthorblockA{\textit{Engineering Department} \\
\textit{University of Massachusetts Boston}\\
i.zografopoulos@umb.edu}
\and
\IEEEauthorblockN{Anca Jurcut}
\IEEEauthorblockA{\textit{School of Computer Science} \\
\textit{University College Dublin}\\
anca.jurcut@ucd.ie}
}

\IEEEaftertitletext{\vspace{-1.5\baselineskip}}

\maketitle

\begin{abstract}
Modern power systems rely on real-time substation communication protocols, such as the Routable Generic Object-Oriented Substation Event (R-GOOSE), for critical control and protection functions. However, these protocols often lack built-in security features and prioritize availability over confidentiality and integrity, making them susceptible to false data injection and denial-of-service attacks. This vulnerability is exacerbated when communications are transmitted over wide-area or public networks. Addressing these cyber threats is essential to comply with current security mandates, including the DOE's defense-in-depth and zero-trust guidelines. This paper introduces QUIC-TRIP, a transparent security methodology for low-latency IP-based industrial communications. By operating at the Open Systems Interconnection (OSI) Transport Layer (Layer 4), the solution encapsulates and protects data flows without affecting the operation of existing protocol endpoints. Baseline echo Round-Trip Time (RTT) results over a Frankfurt-Amsterdam communication path show that the underlying transport-layer proxy used by QUIC-TRIP achieves a lower average RTT than OpenVPN and only 2.88\% higher average RTT than integrated DTLS 1.2, even with DTLS session reuse. We evaluate the resilience of QUIC-TRIP multipath communication under DoS flooding by securing R-GOOSE communications. In these tests, traffic is transparently delivered through three different paths, and QUIC-TRIP forwards the earliest-arriving duplicate while discarding later copies. The framework provides a triple-redundant defense scheme with a measured communication overhead of 32.18\% per enabled proxied path, offering a bounded trade-off between resilience and bandwidth cost for time-critical grid operations.

\end{abstract}

\begin{IEEEkeywords}
Communication, cybersecurity, GOOSE, resilience, substation.
\end{IEEEkeywords}


\section{Introduction} \label{s:intro}

The modernization of power grid infrastructure has resulted in an extensively interconnected cyber-physical system where stability and reliability are critically dependent on high-bandwidth data communication and embedded computing infrastructure \cite{zografopoulos2025cyber}. Substation automation systems rely heavily on protocols, such as the generic object-oriented substation event (GOOSE) that follows the IEC 61850 standard, to facilitate essential, time-critical functions such as asset management, device coordination, protection, and control. However, the certain design characteristics of these protocols can potentially pose substantial cyber challenges. For protocols like GOOSE, the design inherently prioritizes availability and latency minimization over fundamental security traits, such as confidentiality and integrity. Although integrity was traditionally maintained via simple mechanisms, cryptographic methods were omitted. The use of mechanisms like cyclic redundancy checks (CRC), for instance, in related synchrophasor standards like IEEE C37.118, offers weak integrity protection because an intruder can easily recalculate a new CRC after modifying the packet contents. 
Such security deficiencies become a serious issue with the introduction of internet-facing devices that route grid communication over potentially insecure wide-area networks.

Furthermore, the complexity of cyber-physical energy systems (CPES) is exacerbated by the continuous integration of new system elements and the persistence of legacy infrastructure. Legacy devices often suffer from limited computational resources and low communication bandwidth. Many intelligent electronic devices (IEDs) and legacy components, being resource-constrained, make the implementation of rigorous security updates or over-the-air firmware upgrades challenging or infeasible. Although comprehensive security standards have been proposed, such as IEC 62351 for IEC 61850, vendors have yet to adopt them across all products. 
Reliance on vendor-specific development to incorporate security features into their products, 
risks creating additional and potentially unique threat vectors. Consequently, both existing and future infrastructures become susceptible to attacks, such as false data injection attacks (FDIA) and denial-of-service (DoS) \cite{liang2016review}. 

To overcome these limitations and comply with modern mandates, such as the DOE's defense-in-depth and zero-trust guidelines, we propose a novel triple-redundancy security scheme. The primary contributions of this work are as follows:

\begin{itemize}
    \item We develop and open-source\footnote{All source code and supporting data for this study are publicly accessible to ensure reproducibility \cite{reflector}.} a transparent multipath proxying scheme operating at OSI Layer 4 (Transport Layer) to enhance security and provide redundancy for time-critical communication.
    
    \item We analyze how the proposed multipath scheme secures data and provides resilient delivery. This mitigates threats (e.g., FDIA and DoS) targeting critical nodes, removing \textit{single-points-of-failure} associated with centralized industrial gateways, data diodes or virtual private networks (VPNs), which reduce reliability and increase latency. 


    \item We test the solution and quantify the latency and communication overhead from securing R-GOOSE messages. 
\end{itemize}

\section{Background on Communication Protocols}

\begin{table*}[t!]
\footnotesize
\setlength{\tabcolsep}{3pt}
\centering
\caption{Overview of Power Systems Communication Protocol Requirements}
\label{tab:protocols}
\vspace{-2mm}
\begin{tabularx}{\linewidth}{|l|X|>{\centering\arraybackslash}p{0.21\textwidth}|>{\centering\arraybackslash}p{0.215\textwidth}|>{\centering\arraybackslash}p{0.18\textwidth}|}
\hline
\makecell[c]{\textbf{Protocol / Std.}} & \makecell[c]{\textbf{Application}} & \textbf{Limitations} & \textbf{Security} & \textbf{Timing} \\
\hline
\makecell[c]{IEC 61850} & \makecell[c]{Substation automation (GOOSE,SV)} & Complexity, high bandwidth & Limited encryption & GOOSE: $\leq$4ms, SV: $\leq$1ms\\
\hline
\makecell[c]{IEC 60870-5} & SCADA (101: Serial, 104: TCP/IP) & 101: Low rates, 104: IP risks & Vulnerable to injection attacks & 101:1 -- 2s, 104:100ms \\
\hline
IEEE C37.118 &  \makecell[c]{Wide-area monitoring (PMUs)} & Data processing, GPS dependent & GPS spoofing, no encryption & 16 -- 100 ms/frame \\
\hline
\makecell[c]{IEEE 1815\\DNP3} & \makecell[c]{Automation, SCADA} & Latency in large-scale setups & Security extensions available &  \makecell[c]{Reports: 500ms -- 2s \\ Alarms: $\leq$100ms }\\
\hline
\makecell[c]{MODBUS} & \makecell[c]{ICS, Power meters} & Poor error detection, no sync & Vulnerable to command injection & 100ms -- 1s \\
\hline 
\makecell[c]{IEEE 2030.5} & \makecell[c]{DER Internet-based communication} & Computational overheads & TLS 1.2 & Unsuitable for real-time \\
\hline
\end{tabularx}
\vspace{-4mm}
\end{table*}

Securing industrial protocols is challenging, as these protocols are deployed in critical infrastructure systems with long operational lifespans. Protocols are supported by heterogeneous devices from various manufacturers (many of which are resource-constrained), making updates difficult or infeasible. This challenge is further exacerbated in mission-critical deployments, such as electrical systems, where time-sensitive operations and the prohibitive costs of downtime necessitate meticulous planning for upgrades. 

Potential solutions would require that legacy devices can be retrofitted with secure and up-to-date communication protocols or complemented with sidecar hardware without compromising performance. 
CPES often consists of multiple subsystems and industrial processes that demand strict adherence to latency, availability, and performance requirements. Consequently, protocol upgrades must be carefully designed to align with these constraints. In the following subsections, we provide an overview of widely used industrial protocols in electrical systems, their operational and security requirements, and recent advancements aimed at enhancing their security. Finally, we demonstrate how the proposed approach, whilst not a ``silver-bullet solution", can significantly improve the security and reliability of most CPES protocols, without introducing substantial performance overheads or necessitating major investments in infrastructure modernization.

\subsection{Security Analysis of Protocols in Electrical Systems}
Power system monitoring, operation, and control rely on a diverse range of protocols. These are selected based on the specific application domain and technical requirements, such as latency, availability, reliability, and real-time constraints. 
Some of the most commonly used protocols include IEC 61850, IEC 60870, IEEE C37.118, IEEE 1815 (DNP3), MODBUS, and IEEE 2030.5. 
An overview of the characteristics of each protocol and their security oversights is outlined in Table \ref{tab:protocols}.

The IEC 61850 and IEC 60870 families refer to protocol standards that are widely adopted in substations for remote monitoring, automation, and control. IEC 61850 enables real-time communications through the Generic Object-Oriented Substation Event (GOOSE) and Sampled Values (SV) messaging, ensuring low-latency and high-reliability communication. 
Typically, the low latency requirements of IEC 61850 are achieved through wired communication channels, which prohibit the wide adoption of the protocol over longer distances. On the other hand, IEC 60870 can be utilized in both wired and wireless deployments, through standardized communication over TCP/IP networks, for the remote monitoring and control of substations and power plants. Another protocol that prioritizes low latency over security is IEEE C37.118, which is the de-facto standard enabling the transmission of high-resolution time-synchronized data from phasor measurement units (PMU) to control centers. IEEE C37.118 supports wide-area monitoring, event analysis, and stability assessment to improve grid reliability. However, its high-throughput and near real-time transmission requirements have to date prioritized measurement availability and offer limited options for security considerations. \looseness=-1

{DNP3} is extensively used in {SCADA} systems to facilitate secure and reliable communication between industrial, master and outstation devices. It is designed to operate efficiently over long distances and supports robust error checking and time-stamped data for event-driven reporting. Especially its current version, i.e., DNP3-SAv6 \cite{ieee_p1815_sav6}, has been updated to include secure authentication, access control, hash-based message authentication codes (HMAC), and elliptic key cryptography. Similarly, MODBUS was initially developed for industrial automation, facilitating communication between control devices and field equipment such as sensors and actuators. Its simplicity contributed to its widespread adoption, particularly in wired configurations. However, the absence of built-in security mechanisms has raised significant concerns. To address these security vulnerabilities, the SunSpec MODBUS variant was introduced, but its adoption still remains limited \cite{sunspec_modbus}. 

Last, the IEEE 2030.5 is the newest protocol, originally introduced in 2018. IEEE 2030.5 facilitates secure communication between utilities, distributed energy resources (DERs), and customers, playing a crucial role in demand response and grid-edge management. 
The latest IEEE 2030.5-2023 standard incorporates cybersecurity measures, including transport layer security (TLS) for encrypted communications, certificate-based authentication, and role-based access control, ensuring protection against cyber threats and unauthorized access \cite{ieee2030_5_2023}.

The majority of industrial protocols used today have available upgraded versions or extensions that provide valuable security features that were missing in past versions. However, their correct implementation in real deployments is not straightforward, as they rely on secure mechanisms originally designed for general information technology (IT) environments (such as TLS), where operational constraints are more relaxed. Thus, any bolt-on solution involving protocol translation would incur higher latency requirements, more complex middleware, and difficulty in maintaining and scaling secure infrastructure. \looseness=-1

\section{Methodology} \label{s:method}

\begin{figure}[t!]
    \centering
    \includegraphics[width=0.4\textwidth]{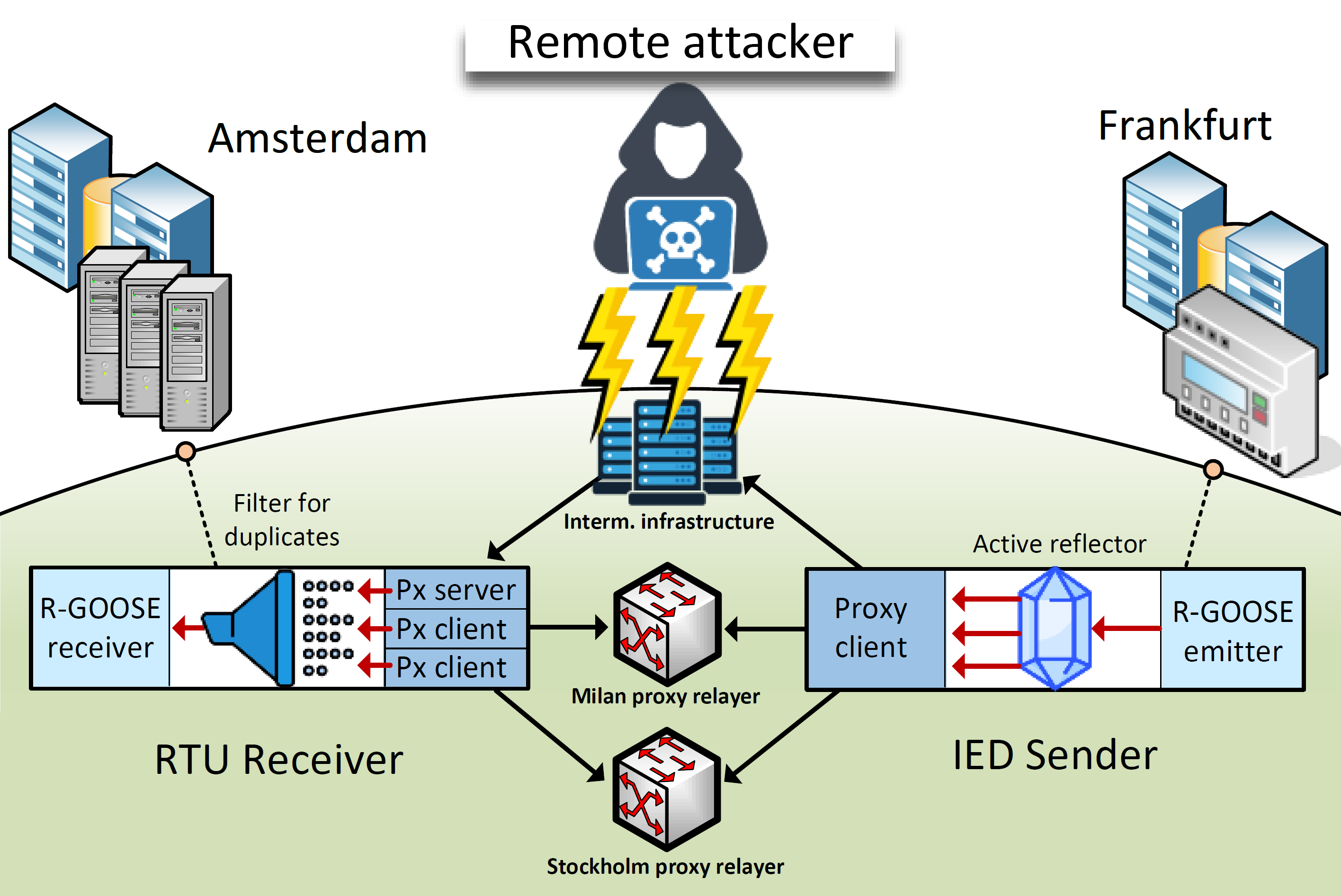}
    \caption{Triple modular redundancy through QUIC proxying (reflected \& filtered communications).}
    \label{fig:attack-reflected}
\end{figure}

This work investigates the communication mechanisms of substation automation systems, emphasizing the operational characteristics and security of GOOSE. 
GOOSE transmits critical protection and control commands, such as trip or interlocking signals. 
GOOSE was originally designed for air-gapped, decentralized local area networks (LANs) using multicast Ethernet with transfer times below 4 ms. 
However, with increasing digitalization, a routable version of GOOSE (R-GOOSE) emerged  to facilitate substations transitioning toward integrated architectures in which human–machine interfaces and gateways aggregate data for transmission to control centers. 
This work aims to develop an application-agnostic, backward-compatible communication scheme that enhances integrity, security, and attack tolerance in modern substations. As demonstrated in Fig. \ref{fig:attack-reflected}, our \textit{QUIC-TRIP} solution establishes a secure \textit{triple modular redundant tunnel} between the network endpoints (e.g., substation and system operator center), providing end-to-end encryption to ensure the \textbf{confidentiality and integrity} of the encapsulated messages while traversing the insecure channels, as well as \textbf{resilience} (i.e., triple modular redundancy) via the multipath transmission.

\subsection{Threat Model} 
\begin{figure}[t!]
    \centering
    \includegraphics[width=0.48\textwidth]{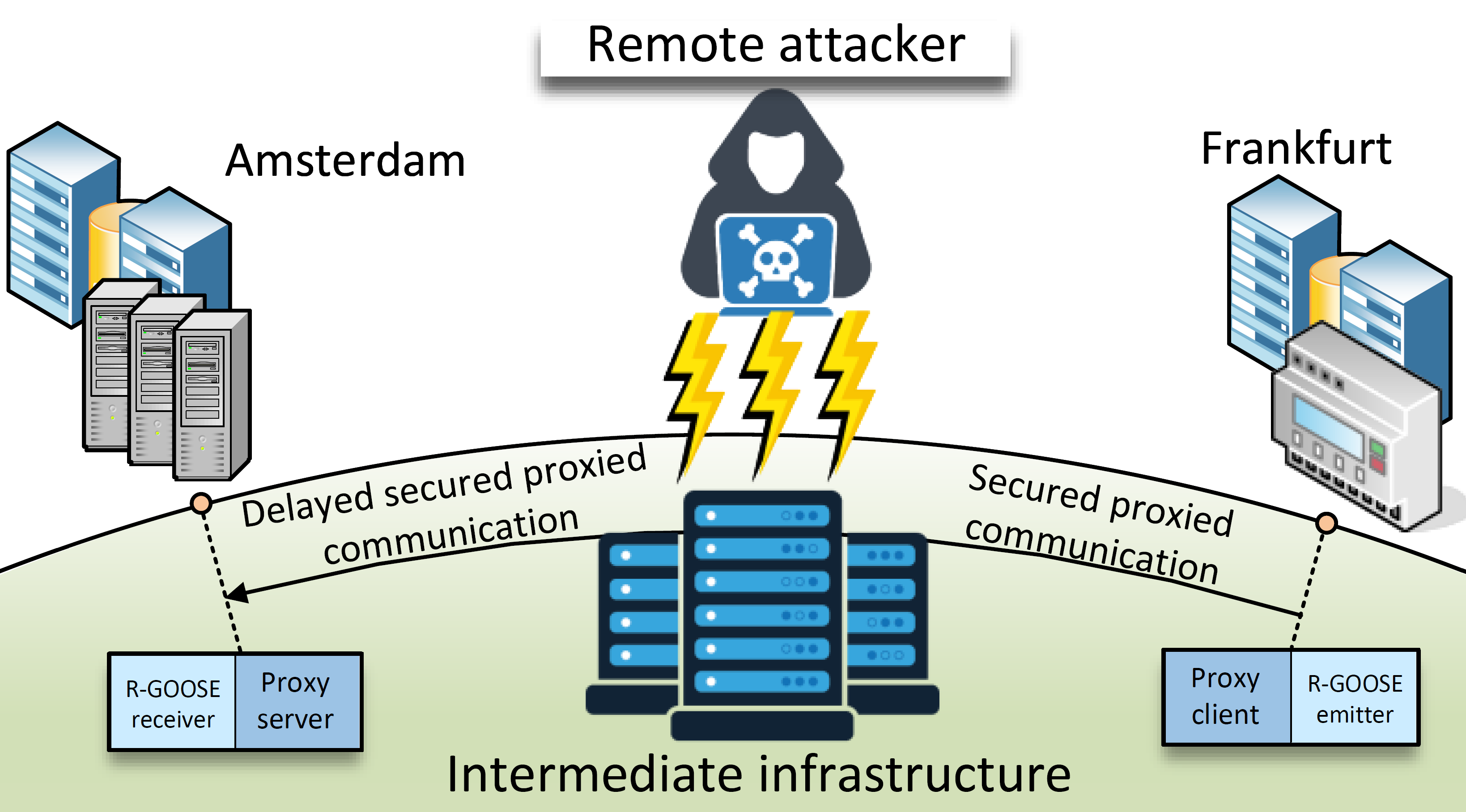}
    \caption{DoS causing communication delays in intermediate infrastructure to proxied communications (non-reflected communications).}
    \label{fig:attack-noreflection}
\end{figure}

As described in \cite{zografopoulos2021cyber}, the \emph{remote attacker} (see Fig. \ref{fig:attack-noreflection}) is modeled as an oblivious adversary having essentially no knowledge of the power system topology, classified as either Class I or Class II, based on the resources and skills required to compromise the CPES. This effort is considered a targeted attack aimed at destabilizing the power grid by obstructing crucial control commands to substations, requiring the attack to be performed iteratively and multiple times to maximize its impact \cite{zografopoulos2022time}. Physical access is not required, as the attack occurs in the cyber domain, 
namely at the network level. The adversarial objectives are FDIA or DoS attacks, which target asset availability by tampering with control commands 
targeting Level 2 assets. The adversaries could leverage the spoofing and resource exhaustion tactics, to induce anomalous incidents and degrade system operations.

\subsection{Baseline latency overhead of major security strategies}
Securing communications from legacy systems can be addressed through three main strategies. Either \textit{i)} by integrating a security technology, combining secure access control with isolated demilitarized zones (VPN-based), by \textit{ii)} using per-device security elements such as sidecar devices,  or \textit{iii)} via stand-alone proxying mechanisms.
Including security in devices has been the standard procedure, mirroring the development processes followed in standard IT systems. However, as discussed in Section \ref{s:intro}, novel solutions are required for resource-constrained legacy systems. 

We compare baseline latency overhead between the three aforementioned secure communication methods, i.e., integrated, VPN-based, and proxy-based. In this evaluation we quantify the round-trip time (RTT) of communications between two different data centers in Europe included in the Linode infrastructure  \cite{linode_infrastructure}. Table \ref{tab:rtt_stats} provides the RTT between Linodes in Frankfurt and the Amsterdam, when performing user datagram protocol (UDP) echo tests secured using \textit{i)} Datagram Transport Layer Security (DTLS), \textit{ii)} OpenVPN, and \textit{iii)} a QUIC-based proxy based on SSH3 \cite{Michel2023SSH3}. To achieve a fair comparison with DTLS, we developed a custom echo server/client program to fine-tune communications and measure RTTs for each technology analyzed. This enables us to use the same UDP client/server in either a DTLS-secure mode for integrated security evaluation, or in an unsecured mode for VPN and proxy-based solutions.

\begin{table}[t!]
\centering
\footnotesize
\caption{RTT baseline statistics comparison between \\ Frankfurt and Amsterdam Linodes}
\label{tab:rtt_stats}
\vspace{-1mm}
\setlength{\tabcolsep}{3pt}
\begin{tabular}{lccc}
\toprule
\textbf{Metric} &
\begin{tabular}[c]{@{}c@{}}\textbf{Integrated DTLS 1.2*}\\(OpenSSL 3.0.13)\end{tabular} &
\begin{tabular}[c]{@{}c@{}}\textbf{OpenVPN}\\(2.6.14)\end{tabular} &
\begin{tabular}[c]{@{}c@{}}\textbf{SSH3 Proxy}\\(QUIC--TLS~1.3)\end{tabular} \\
\midrule
RTT Avg. (ms)         & 6.25 & 6.97 & 6.43 \\
RTT $\sigma$ (ms)     & 0.11 & 0.13 & 0.13 \\
\midrule
RTT Avg. increase     & ---  & 11.52\% & \textbf{2.88\%} \\
RTT $\sigma$ increase & ---  & 18.18\% & 18.18\% \\
\bottomrule
\end{tabular}

\vspace{1mm}
\begin{minipage}{0.98\linewidth}
\footnotesize
*To achieve these results, each DTLS datagram reuses the same session.
\end{minipage}
\vspace{-4mm}
\end{table}


In industrial environments with legacy systems, VPN-based solutions are the standard for securing communication that must leave the operational technology (OT) network. They operate at OSI level 3 (Network Layer) and are expected to be faster than external solutions operating at higher levels. Still, the efficiency of the QUIC protocol enables the development of streamlined communication technologies, such as SSH3 \cite{Michel2023SSH3}, that allow developing flexible proxying capabilities with minimum average latency increase (i.e., 2.88\%)\cite{CCNC2026letsgetquic}. \looseness=-1

Although integrated security using DTLS performs slightly better, it requires that \textit{the same session be maintained} throughout the tests. Since establishing a session can take several seconds, TLS 1.3 can be employed to speed the process thanks to 0-RTT handshake session resumption. However, this special feature is generally avoided for security reasons unless “\textit{an explicit specification exists for the application protocol in question to clarify when 0-RTT is appropriate and secure},” as is the case with HTTP/3 \cite{rfc9325}. 

\subsection{QUIC for Low-latency and Multi-path Redundancy}
Redundancy for critical communications has been used systematically, with dedicated hardware and protocols that increase reliability and ensure operation below maximum latency limits. However, they are not readily applicable to arbitrary deployments, as such redundancy requires planning during design phases. To address this issue, Saldana et al. \cite{Jaldana2023} proposed an application-agnostic communication method that works independently of the application protocol by replicating critical datagrams multiple times, which are later filtered at the destination. This approach mitigates the disruptive effects of packet loss in unreliable environments while preserving low latency. However, it comes at the expense of additional bandwidth usage for actively replicating datagrams in blasts.

Building on this idea, we explore using multipath to send these copies, further mitigating potential communication effects by introducing \textit{multipath diversity}. This is achieved in a controlled manner by exploiting a proxy-based solution that allows planning the path that duplicated datagrams should follow in advance \cite{cmxsafe}. In our scheme, application-agnosticity is also preserved, i.e.,  datagrams are proxied at an OSI level below the application level. Furthermore, we provide security guarantees and the ability to duplicate datagrams through pre-established and secure communication paths. Our approach leverages an SSH3-based standalone proxy solution \cite{CCNC2026letsgetquic}, which provides finer-grained access control compared to VPNs or virtual LANs (up to the process level \cite{cmxsafe}), but without having to integrate into the application layer. In this regard, the 
QUIC working group is developing an extension for QUIC that will support multipath natively \cite{I-D-quic-multipath-16}. However, it is not designed to constantly replicate datagrams across all available paths.  \looseness=-1 

Thus, in this work, we propose a low-latency multipath redundancy scheme using an active reflector on the sender side and a filter on the receiver side. Three paths are used as a practical minimum for path diversity and triple-modular redundancy, while keeping bandwidth overhead bounded. The approach is not limited to three paths and can be generalized to \(N\) paths, with the corresponding traffic overhead scaling as discussed in Section~\ref{sec:overhead}. This approach leverages a  proxying layer, which allows a precise mapping of features into independent components. Specifically,  \textit{i)} the application generates communication datagrams using the R-GOOSE protocol every 10ms~\cite{10066414}, \textit{ii)} the reflector copies each arriving UDP datagram and appends a 4-byte label that includes a datagram number shared across each set of replicated datagrams. Then, \textit{iii)} each duplicated datagram is securely proxied through different end-to-end secured communication paths. 
At the destination, \textit{iv)} a filter maintains a counter of the set of received replicated datagrams that match the label of incoming duplicated datagrams. These datagrams are securely delivered to the filter by the stand-alone secure proxy, and the filter provides the destination-side application with only the first-arriving datagram of each set of copies (i.e., modular redundancy). \looseness=-1

\subsection{QUIC-based Testbed Setup}
The Linode-based testbed is not intended to reproduce substation LAN latency, but to emulate wide-area IP transport between substation and control-center endpoints under controlled congestion. This enables repeatable evaluation of path diversity and DoS-induced delay without disrupting operational infrastructure. As shown in Fig.~\ref{fig:attack-reflected}, alternative routes are selected across different datacenters to maximize resilience. We place the IED sender in the Frankfurt datacenter and consider three routes to the remote terminal unit (RTU) receiver in Amsterdam.


Route 1 follows the direct path through the existing intermediate infrastructure. A Linode is placed as a standard network router (OSI Layer 3) through which all traffic to Amsterdam must pass, allowing DoS progression to be modeled by targeting this router. Routes 2 and 3 are alternative paths through the Stockholm and Milan datacenters, respectively. In each location, a Linode acts as a proxy relayer (OSI Layer 4 gateway), enabling the IED sender and RTU receiver to establish proxying sessions for secure end-to-end communication \cite{cmxsafe}.

 \begin{figure}[t!]
    \centering
    \includegraphics[width=0.5\textwidth]{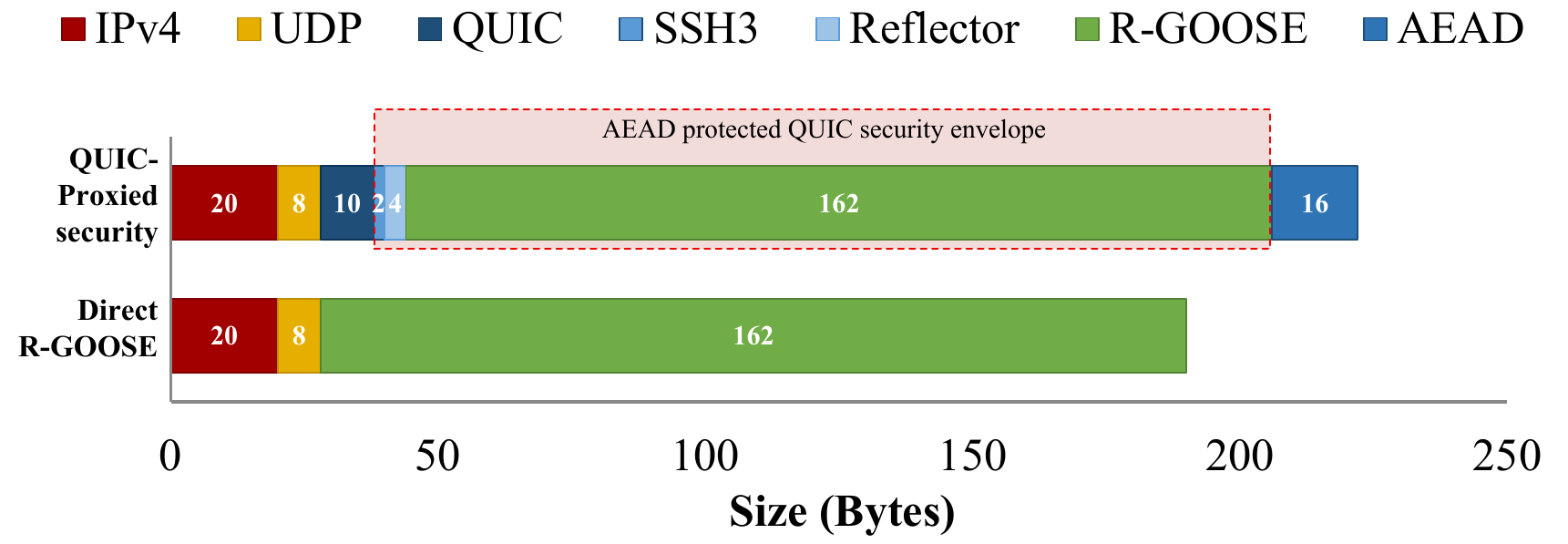}
    \caption{Packet structure comparison between unsecured and  QUIC-TRIP secured proxied R-GOOSE communication of a 162-byte datagram.}
    \vspace{-5mm}
    \label{fig:packet-structure}
\end{figure}

The QUIC-based proxy mechanism in Fig.~\ref{fig:attack-reflected} relies on SSH3 over QUIC for transport and security~\cite{CCNC2026letsgetquic}. As shown in Fig.~\ref{fig:packet-structure}, the SSH3 forwarding metadata, reflector tag, and R-GOOSE datagram are carried inside the QUIC-protected payload. QUIC uses TLS~1.3 for connection establishment, authentication, and packet-protection key derivation, while encryption and authentication are applied by QUIC rather than conventional TLS records. The negotiated TLS~1.3 cipher suite selects the Authenticated Encryption with Associated Data (AEAD) algorithm, such as AES-GCM or ChaCha20-Poly1305, which adds a 16-byte tag binding the encrypted payload to the QUIC header context. Packets failing AEAD verification are discarded.\looseness=-1

To evaluate the resilience of the proposed secure multipath proxying mechanism, we use a standard DoS flooding attack aimed at saturating network capacity. The first scenario targets only the intermediate infrastructure (Fig.~\ref{fig:attack-noreflection}), while the second also targets the proxy relayers in the alternative datacenters (Fig.~\ref{fig:attack-reflected}). Because Linode is a commercial Infrastructure-as-a-Service (IaaS) provider, unrestricted flooding against production resources is neither appropriate nor permissible. Therefore, the router and proxy-relayer links are limited to 5~Mbps, allowing controlled stress testing without disrupting provider infrastructure or other tenants. In each attack, we evaluate the following two configurations simultaneously:

 \textit{1) Non-reflected communications (Fig. \ref{fig:attack-noreflection}):} R-GOOSE emitter is secured through proxying \textit{without reflection} through the intermediate networking infrastructure utilizing the default fastest route.
 
 \textit{2) Reflected \& Filtered communications (Fig. \ref{fig:attack-reflected}):} R-GOOSE emitter is secured through our proposed reflection mechanism through three routes (intermediate infrastructure, Milan path, and Stockholm path). 
 
 The fastest route (Route 1) uses the existing intermediate infrastructure where the DoS-targeted router is located. This router is shared by both configurations (since one of the redundant paths of the second configuration also flows through the intermediate infrastructure). The other two paths flow through secure communication paths established with proxy relayers running on Linodes, placed in Milan and Stockholm datacenters  \cite{cmxsafe}. These paths incur higher latency, but provide redundant communication paths decoupled from the intermediate network infrastructure used in the default fastest communication path.

\begin{table}[t]
\centering
\caption{Breakdown of the measured forwarding overhead when compared to plain R-GOOSE unsecured traffic}
\label{tab:rgoose-overhead-breakdown}
\begin{tabular}{l r}
\hline
\textbf{Component} & \textbf{Overhead} \\
\hline
Per-datagram expansion & 16.86\% \\
Extra client-to-server control traffic & 0.68\% \\
Server-to-client responses & 14.64\% \\
\hline
\textbf{Total measured overhead} & \textbf{32.18\%} \\
\hline
\end{tabular}
\vspace{-4mm}
\end{table}

\section{Results and Discussion} \label{s:results}
Building on the testbed and attack scenarios described above, this section evaluates QUIC-TRIP overhead and resilience. It quantifies QUIC/SSH3 proxying and reflection costs through theoretical per-datagram expansion and measured bidirectional overhead, before assessing DoS-induced latency against a non-reflected proxied baseline. \looseness=-1

\subsection{Communication Overhead Analysis}
\label{sec:overhead}
From the packet structure in Fig.~\ref{fig:packet-structure}, the forwarding overhead is estimated at 32 bytes per forwarded R-GOOSE datagram. In the traces, original IPv4-layer packets were about 189-191 bytes, compared with 221-223 bytes for the corresponding QUIC/SSH3 data-only packets, giving a theoretical overhead of 16.8-16.9\% per datagram. This accounts for the reflector label, SSH3 forwarding metadata, and QUIC protection, including AEAD, but excludes PINGs, ACKs, control packets, and server-to-client responses needed to maintain the QUIC/SSH3 session. Therefore, we also measured attack-free R-GOOSE traffic over 10 minutes. For 162-byte R-GOOSE datagrams, the complete added overhead was 32.18\%, as described in Table~\ref{tab:rgoose-overhead-breakdown}.
\looseness=-1

The overhead of one proxied path is incurred on every redundant path. Therefore, relative to a single original R-GOOSE communication, the total overhead for \(N\) paths is:
\[
O_{\mathrm{total}}(N)=N\left(1+O_{\mathrm{path}}\right)-1,
\]
where \(O_{\mathrm{path}}\) is the overhead ratio of one proxied path and \(N\) is the total number of communication paths.

\subsection{Resilience Under DoS Flooding Attacks}

\begin{figure}[t!]
    \centering
    \includegraphics[width=0.5\textwidth]{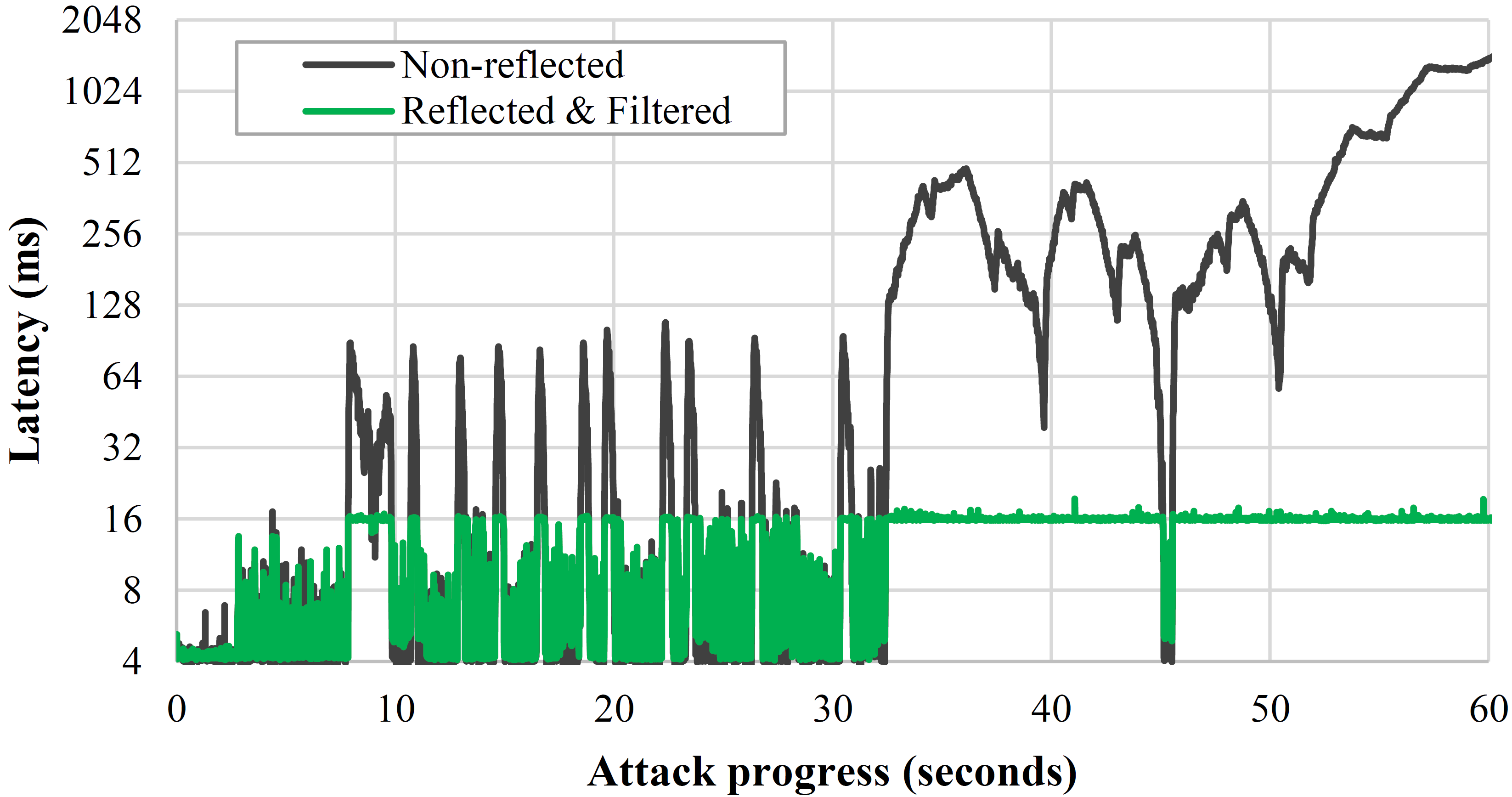}
    \caption{Attack trace to the intermediate infrastructure. Reflected traffic avoids the affected datacenter (Frankfurt). The filter processes the earliest-arriving copied datagrams (Milan datagrams).}
    \vspace{-4mm}
    \label{fig:attack}
\end{figure}

The first flooding attack targets the router in the intermediate infrastructure used by both configurations. As shown in Fig.~\ref{fig:attack}, the latency impact has three stages. First, early attack packets ($0$--$8$s) reach the attacked router, through which traffic arrives at the destination RTU in both the non-reflected configuration and one reflected path. This causes a moderate latency increase in both cases, between $10$-$16$ ms. Then, flooding bursts accumulate in the router buffers, producing short-lived spikes of up to $100$ ms in the non-reflected configuration. By contrast, the reflected-and-filtered configuration discards higher-latency datagrams when lower-latency counterparts have already arrived through alternate routes. In this example, the selected route traverses Milan, with a baseline end-to-end latency of $\sim17$ ms, while the Stockholm alternate-failover route is $\sim22$ ms. \looseness=-1

In the reflected-and-filtered configuration, QUIC provides security and ordered delivery over each proxied path, simplifying the design of the filter and reflector. This enables a highly optimized standalone implementation with minimal latency overhead ($100$--$300$~\textmu{}s). During an attack, the filter forwards only the first-arriving datagram of each duplicate set to the application. It therefore does not need to handle ordering, authentication, or security checks, which are provided by QUIC, TLS~1.3, and the proxying layer.

However, this comes with increased bandwidth due to traffic replication  as seen in Table \ref{tab:rgoose-overhead-breakdown}. 

\begin{figure}[t!]
    \centering
    \includegraphics[width=0.5\textwidth]{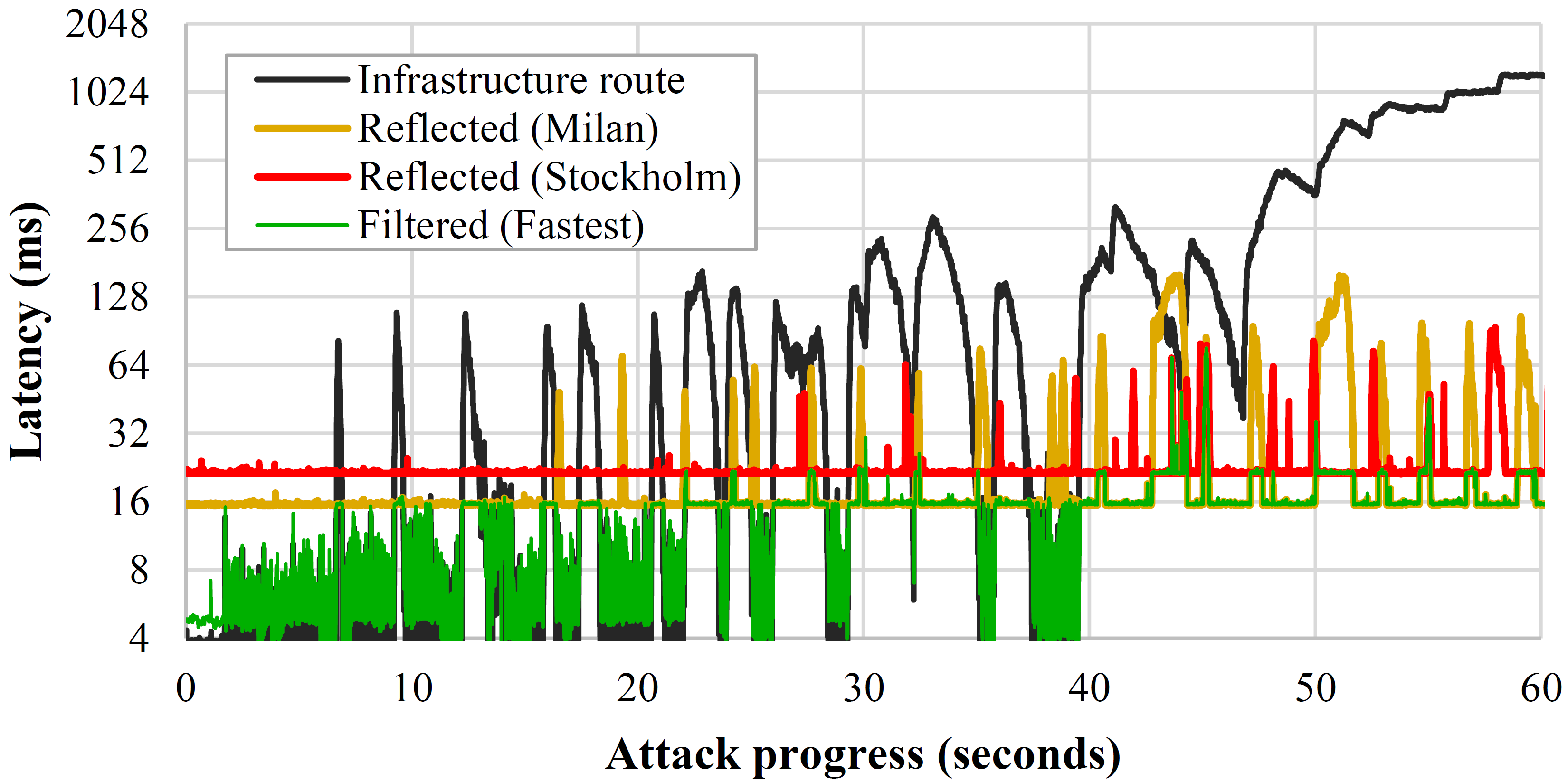}
    \caption{Attack trace to the intermediate infrastructure and alternative paths. Non-reflected traffic and reflected traffic through the infrastructure overlaps. Filter minimizes the effects by forwarding to the application only the fastest datagram of each reflected set.}
    \vspace{-3mm}
    \label{fig:attack2}
\end{figure}

The second attack extends its influence to the proxy relayers used by the securely proxied reflected datagrams. To facilitate trace analysis, flooding toward Milan's proxy relayer is delayed by $10$~s, while flooding toward Stockholm's proxy relayer is delayed by $20$~s, as shown in Fig.~\ref{fig:attack2}. The reflected and filtered configuration mitigates the attack by forwarding reflected datagrams that arrive faster to the destination and discarding later duplicates. If the attack continues and intensifies simultaneously across all paths, eventual increases in latency become unavoidable. Still, the multipath scheme allows new paths to be established in the background through different routes, enabling a congested proxied path to relay the reflected datagram stream to a newly established secure path.\looseness=-1

The presented multipath communication scheme provides a systematic way to control latency under attacks affecting intermediate networking infrastructures. Furthermore, it enables trading additional communication bandwidth for packet delivery assurance, which is viable in mission-critical systems with sporadic communications such as R-GOOSE. With \textit{QUIC-TRIP}, it is possible to balance attack-mitigation costs against the baseline latency of the available detours. Likewise, the degree of reliability depends on the number of alternative paths, the diversity of infrastructures/datacenters involved, and the allocated bandwidth, as the \textit{QUIC-TRIP} number of reflected paths can be easily extended to improve resilience. 
\looseness=-1






\section{Conclusions}

This paper introduces QUIC-TRIP, a triple-redundant reflector and filter methodology that enhances substation communication security and data-delivery resilience under adverse cyber conditions. In a UDP echo RTT benchmark over the Frankfurt-Amsterdam Linode path, the SSH3/QUIC proxy showed lower average RTT than OpenVPN and was only $2.88\%$ higher than integrated DTLS~1.2 with session reuse. This low RTT overhead motivated its use as a feasible substrate for QUIC-TRIP, enabling transparent multipath delivery through active reflection and filtering. In the R-GOOSE testbed, the scheme reduced the latency impact of DoS flooding,  relative to the non-reflected proxied baseline, by forwarding the earliest-arriving duplicate and discarding later copies. This resilience comes with a measured communication overhead of $32.18\%$ per enabled proxied path when securing 162-byte R-GOOSE datagrams. Future work will involve real-time co-simulation and deployment to assess the impact of communication delays on power system stability.

\section*{Acknowledgment}
This work is funded in part by the National Science Foundation (NSF) Award Number \#2501975. This work is supported in part by the European Union’s Horizon Europe research and innovation programme under the Marie Skłodowska-Curie Actions grant agreement No. 101149974 (Project CMXsafe).


\bibliographystyle{IEEEtran}
\bibliography{biblio}

\end{document}